\documentclass[english,twocolumn]{aastex62}
\usepackage[T1]{fontenc}
\usepackage{float}
\usepackage{amstext}
\usepackage{amssymb}
\usepackage{amsmath}
\usepackage{graphicx}
\usepackage{color}
\usepackage{ulem}
\usepackage{natbib}
\usepackage[caption=false]{subfig}

\makeatletter
\shorttitle{}
\shortauthors{}

\usepackage{multirow}
\usepackage{rotating}

\makeatother

\usepackage{babel}
\begin{document}

\title{The Impact of Initial Composition on Massive Star Evolution and Nucleosynthesis}

\author{Christopher West}
\affil{Department of Physics and Astronomy, Carleton College, Northfield, MN 55057, USA}
\affil{Department of Physics and Astronomy, Macalester College, Saint Paul, MN 55105, USA}
\affil{School of Physics and Astronomy, University of Minnesota, Minneapolis, MN 55455, USA}
\affil{Minnesota Institute for Astrophysics}
\affil{Joint Institute for Nuclear Astrophysics, Notre Dame, IN 46556, USA}
\email{cwest@carleton.edu, christopher.west287@gmail.com}

\author{Alexander Heger}
\affil{School of Physics and Astronomy, Monash University, Vic 3800, Australia}
\affil{Joint Institute for Nuclear Astrophysics, Michigan State University, MI, USA}
\email{alexander.heger@monash.edu}

\author[0000-0002-9986-8816]{Benoit C\^ot\'e}
\affil{Konkoly Observatory, Research Centre for Astronomy and Earth Sciences, Hungarian Academy of Sciences, Konkoly Thege Miklos ut 15-17, H-1121 Budapest, Hungary}
\affil{ELTE E\"otv\"os Lor\'and University, Institute of Physics, Budapest, Hungary}
\affil{Joint Institute for Nuclear Astrophysics - Center for the Evolution of the
Elements, USA}
\affil{NuGrid Collaboration, \url{http://nugridstars.org}}

\author{Lev Serxner}
\affil{Department of Physics and Astronomy, Macalester College, Saint Paul, MN 55105, USA}
\email{lssgm1@gmail.com}

\author{Haoxuan Sun}
\affil{Department of Physics and Astronomy, Macalester College, Saint Paul, MN 55105, USA}
\email{hsun@macalester.edu}

\begin{abstract}
We study the sensitivity of presupernova evolution and supernova nucleosynthesis yields of massive stars to variations of the initial composition. We use the solar abundances from Lodders (2009), and compute two different initial stellar compositions: i) scaled solar abundances, and ii) the isotopic galactic chemical history model (GCH) developed by West and Heger (2013b).  We run a grid of models using the KEPLER stellar evolution code, with 7 initial stellar masses, 12 initial metallicities, and two for each scaling method to explore the effects on nucleosynthesis over a metallicity range of $-4.0\leq[Z]\leq+0.3$. We find that the compositions from the GCH model better reproduce the weak \emph{s}-process peak than the scaled solar models. The model yields are then used in the OMEGA Galactic Chemical Evolution (GCE) code to assess this result further. We find that initial abundances used in computing stellar structure have more of an impact on GCE results than initial abundances used in the burn network, with the GCH model again being favored when compared to observations. Lastly, a machine learning algorithm was used to verify the free parameter values of the GCH model, which were previously found by West and Heger (2013b) using a stochastic fitting process. The updated model is provided as an accessible tool for further nucleosynthesis studies.
\end{abstract}


\section{Introduction}

The final nucleosynthesis products from massive star simulations are dependent on the initial isotopic composition. As discussed in \citet{West2013b}, for many stars the initial composition is important for neutron capture reactions on initial metals during hydrostatic burning phases. Also, the weak \emph{s}-process yields are affected by the initial CNO abundance, which provides the neutron source, and the initial Fe abundance, which provides the seeds \citep{Pignatari2010}. Moreover, the distribution of the stellar abundances can affect the opacity of the star (e.g., the \citealp{OpacityProject}), which impacts the structure, mass, angular momentum loss, and ultimately the late stellar evolution. Additionally, the $\gamma$-process relies on \emph{s}- and \emph{r}-process metals as seeds \citep{Woosley1978,Rayet1990,Rauscher2002,Arnould2003a}, hence differences in these seed abundance can change the final $\gamma$-process yields. Finally, a large fraction of the initial mass will return to the interstellar medium (ISM) only partially processed by hydrostatic burning and explosive nucleosynthesis, and thus the initial composition directly affects the final abundances.

Ideally, the initial composition would come directly from observations, but unfortunately non-solar isotopic data is scarce. A second choice is to use galactic chemical evolution (GCE) models to provide the needed abundances. This type of modeling, however, relies on stellar simulations as inputs themselves (which require initial compositions), and often do not address all isotopes and astrophysical processes (e.g., \citealp{Timmes1995,Kobayashi2006}). Another challenge in GCE modeling is poorly constrained dynamical quantities, such as inflow/outflow, mixing, and feedback, however, more recent studies have made progress addressing stellar feedback (e.g., \citealp{Emerick2018}, analyzing the effects/trade-offs between star formation efficiency, initial mass function, Type Ia delay time distribution, etc.,  (e.g., \citealp{Andrews2017}, refining data/input needs to delineate GCE results (e.g., \citealp{Cote2016,Cote2017}), and identifying model discrepancies (e.g., \citealp{Mishenina2017}). Despite ongoing
limitations, GCE models have been useful for obtaining initial compositions for massive star simulation inputs (e.g., \citealp{Woosley1995}), although differences in results due to estimated initial abundances have not been well studied.

A common estimation for isotopic abundances used in stellar models (independent of GCE modeling) is to linearly interpolate abundances as a function of metallicity between the known endpoints of their solar and big bang nucleosynthesis (BBN) values. This method effectively assumes that all metals are made in primary processes (i.e., directly from H and He), since primary processes produce abundances that scale linearly in metallicity (since the Galaxy's average metallicity increases monotonically with time, the primary process production rate is time-independent (e.g., \citealp{Timmes1995,Prochaska2003,Bensby2004,Li2008}), however, such relations are dependent on selection effects \citep{Edvardsson1993,Feltzing2001}. This linear interpolation approach is reasonable given that the majority of metals by mass are indeed of primary origin (e.g., light $\alpha$-isotopes and iron-peak isotopes). An obvious error in this method, however, is that isotopes made in secondary processes (i.e., from pre-existing metals) are approximated poorly, since a secondary process event should produce abundances that scale quadratically in metallicity. (secondary processes produce isotopes at a rate proportional to abundances from primary processes). The relations discussed above hold for individual astrophysical events, and ignore abundances produced from a "mixed" primary/secondary history that would more closely describe an averaged Galactic ISM. Nevertheless, by using a linear interpolation method, secondary processes are likely \emph{over-estimated} at sub-solar metallicities and possibly \emph{under-estimated} at super-solar metallicities. 

Moreover, this method also incorrectly over-approximates Type Ia SNe yields at low metallicities $\left([Z]\lesssim-1\right)$. Although Type Ia SNe are primary events, they cannot contribute to the ISM until sufficient time elapses for white dwarfs to evolve from low mass stars and then accrete enough material to explode (assuming an accretion progenitor model, e.g., \citealp{Ruiter2010}). Regardless of progenitor model, Type Ia events should experience a time delay in their ISM contributions not accounted for with a simple linear interpolation, which implicitly assumes Type Ia contributions exist at all metallicities. It is poorly known in detail how these errors impact the final nucleosynthesis. Lastly, uncertainty of the dominant Ia progenitor model further frustrates the analysis. Type Ia SNe simulations show explosion yield differences in sub Fe-peak nucleosynthesis between deflagration, denotation, and remnant models (e.g., \citealp{Kirby2019,Foley2013}, and references therein). In modeling the chemical histories, this may impact the metallicity range where Type Ia contributions become significant (Type Ia onset), and the relative contributions of both Type Ia and massive star nucleosynthesis to the solar abundances \citep{West2013b}.

In this work, we investigate the dependence of the final nucleosynthesis of massive stars on changes in the initial composition, and its consequence in GCE modeling. The initial composition for the stellar models is computed in two ways. The first is the scaled solar abundances discussed above, where abundances are linearly interpolated between their solar values (taken from \citealp{Lodders2010}) and BBN values (taken from \citealp{Fields2002}). In the second method, the initial composition is computed using the isotopic galactic chemical history (GCH) model developed by \citet{West2013b}. In the latter, secondary processes and Type Ia SNe abundances are treated separately from primary processes, which offers a correction to linearly interpolating all abundances. The purpose here is to assess the impact of these corrections on the stellar models, and ascertain whether they result in nucleosynthesis that better matches both observations, and our understanding of the astrophysical processes in massive stars that produce them. 

In particular, we study the impact on the weak \emph{s}-process, which is a slow (relative to the beta decay rate, e.g., \citealp{B2FH}) neutron capture process that occurs in massive stars, and is responsible for isotopic production in the mass range $70\lesssim A \lesssim 100$. For details and discussion of this process see, for e.g., \citet{Pignatari2010,Heil2007}. Note that recent investigation into the $^{17}$O$(\alpha,n)^{20}$Ne and $^{17}$O$(\alpha,\gamma)^{21}$Ne rates suggest production may exceed $A>100$ \citep{Frost2022}, but such consideration is beyond the scope of the present work.

For recent experimental efforts in measuring the Maxwellian-Averaged cross-sections that approach the energy regime for the weak \emph{s}-process, see \citet{Erbacher2023}. For a new measurement of the $^{72}$Ge$(n,\gamma)$ cross-section see \citet{Dietz2023}. See also \citet{Millman2023} for recent GCE calculations using an improved $^{22}$Ne$(\alpha,n)^{25}$Mg rate. 

This paper has the following outline: In \S\ref{sec:Method} we describe the stellar models and procedures used in the analysis, in \S\ref{sec:Results} we present the results of the stellar models, differences in yields compared to observations, and discuss the implications for GCE models. Finally, in \S\ref{sec:Conclusion} we make concluding remarks.

\section{Stellar Models and Numerical Method}
\label{sec:Method}

All models were computed using the KEPLER code, a time-implicit one-dimensional hydrodynamics package for stellar evolution \citep{Weaver1978,Rauscher2002}. We used $8$ different initial stellar masses $M/\mathrm{M}_{\text{\ensuremath{\odot}}}=13,15,17,20,22,25,27,$ and $30$. For each initial mass, a grid of $13$ models were run with the following initial metallicities, $\mathrm{[Z]}=+0.2$, $+0.1$, $0.0$, $-0.2$, $-0.4$, $-0.6$, $-0.8$, $-1.0$, $-1.5$, $-2.0$, $-2.5$, $-3.0$, and $-4.0$, where $[Z]\equiv\mathrm{log\left(Z/Z_{\odot}\right)}$. This entire grid of $104$ models was run four times, using different combinations of initial compositions, explained further below.

\subsection{Stellar Evolution and Nucleosynthesis Calculations}

As discussed in \citet{West2013a}, all stellar models were first evolved through hydrostatic burning until the Fe-core collapsed, and an inward velocity of $10^{8}\,\mathrm{cm\, s^{-1}}$ was reached. The explosion mechanism in KEPLER for the resulting supernova is modeled as a mechanical piston that imparts an acceleration at constant Lagrangian mass coordinate to provide the desired total kinetic energy of the ejecta, taken in these models to be $1.2$\,B ($\mathrm{1\,B}=10^{51}\,\mathrm{erg}$) at $1$ year after the explosion. For details on the parametrization of the explosion used in KEPLER see \citet{Woosley2007}, and references therein. For details on the treatment of convection and mixing see \citet{Woosley1988b} and \citet{Woosley2002}, and a discussion of the mass cut is given in \citet{Tur2007,Heger2010}.

The initial composition of the models affects the final stellar yields in two ways: (a) it directly impacts the nucleosynthesis by constraining the abundances available for nuclear reactions, and (b) it affects the stellar structure and subsequent evolution. The latter does not immediately impact the nucleosynthesis, but it can affect the yields by changing the conditions that permit certain reactions to take place (e.g., \citealp{West2013a}). We distinguish between these two effects in the present work by computing four complete sets of stellar models using scaled-solar abundances and isotopic galactic chemical history (GCH) model abundances from \cite{West2013b} for the initial compositions used in determining both the nucleosynthesis and the structure (see, Table\,\ref{Table1}).

\begin{table}[th]
\centering
\begin{tabular}{|c|c|c|c|}
\cline{3-4}
\multicolumn{2}{c|}{} & \multicolumn{2}{|c|}{\phantom{\huge g}Nucleosynthesis\phantom{\huge g}} \\
\cline{3-4}
\multicolumn{2}{c|}{} & \phantom{\huge g}$\!\!\!\!\!$Galactic Model$\!\!\!\!\!$\phantom{\huge g} & \phantom{\huge g}$\!\!\!\!\!$Scaled Solar$\!\!\!\!\!$\phantom{\huge g} \\
\hline
\multirow{7}{*}{\rotatebox{90}{Structure}}
&&&\\
& Galactic Model & GG & SG \\
&&&\\
\cline{2-4}
&&&\\
& Scaled solar & GS & SS \\
&&&\\
\hline
\end{tabular}
\caption{Initial compositions of the four sets of models. Abbreviation format states composition used for nucleosynthesis (first) and stellar structure (second).\label{Table1}}
\end{table}

\subsection{Initial Composition}

The initial compositions from the scaled-solar and GCH model are identical at $[Z]=0$, where both are the solar abundance pattern \citep{Lodders2010}. We do not address the effects of different solar abundance compositions in the present work. When the metallicity decreases, the ratio of the GCH model over scaled-solar for many of the isotopic abundances becomes increasingly negative. This occurs because the GCH model's compositions scale secondary and Type Ia SNe isotopic abundances at a larger rate than the linearly-scaled, solar abundances. The \emph{s}-process peak isotopes (such as \textsuperscript{134,136}Ba), $\gamma$-process isotopes (such as \textsuperscript{156,158}Dy), and Fe-peak isotopic abundances all display this effect, which is greatest at the lowest initial metallicity considered ($[Z]=-4$). At super-solar metallicities the reverse is true. Here, the increased scaling rate of \emph{s}-process and $\gamma$-process isotopes pushes this ratio more positive, however, the treatment of Type Ia SNe by \citet{West2013b} (a $\tanh$ function scaling) flattens the scaling rate for these relatively abundant isotopes upon reaching solar metallicities so as to not over-produce them relative to observations. The third \emph{s}-process peak isotopes (\textsuperscript{204,206,207,208}Pb and \textsuperscript{209}Bi) depart from this trend. These isotopes are presently understood to be a result from the main \emph{s}-process at low metallicities (see, for e.g., \citealp{Bisterzo2010}), hence the ratio is positive for these isotopes at low metallicities, and negative at super-solar metallicities in accordance with their believed mechanism for production (discussed further in \citealp{West2013b}). A table showing numerical values for both initial compositions for $[Z]=-1,-3$ is provided by \citet{West2013b}.

\section{Results \& Discussion}
\label{sec:Results}

At solar metallicity, all four model sets have the same composition and so produce identical results. One interesting question, however, is which sub-solar model set produces elemental ratios that agree with expected trends. We proceed by addressing weak \emph{s}-process results between the sets.

\subsection{Fe Production and Neutron-to-Seed Ratio}

\begin{figure*}[t]
\includegraphics[angle=180,width=0.48\textwidth]{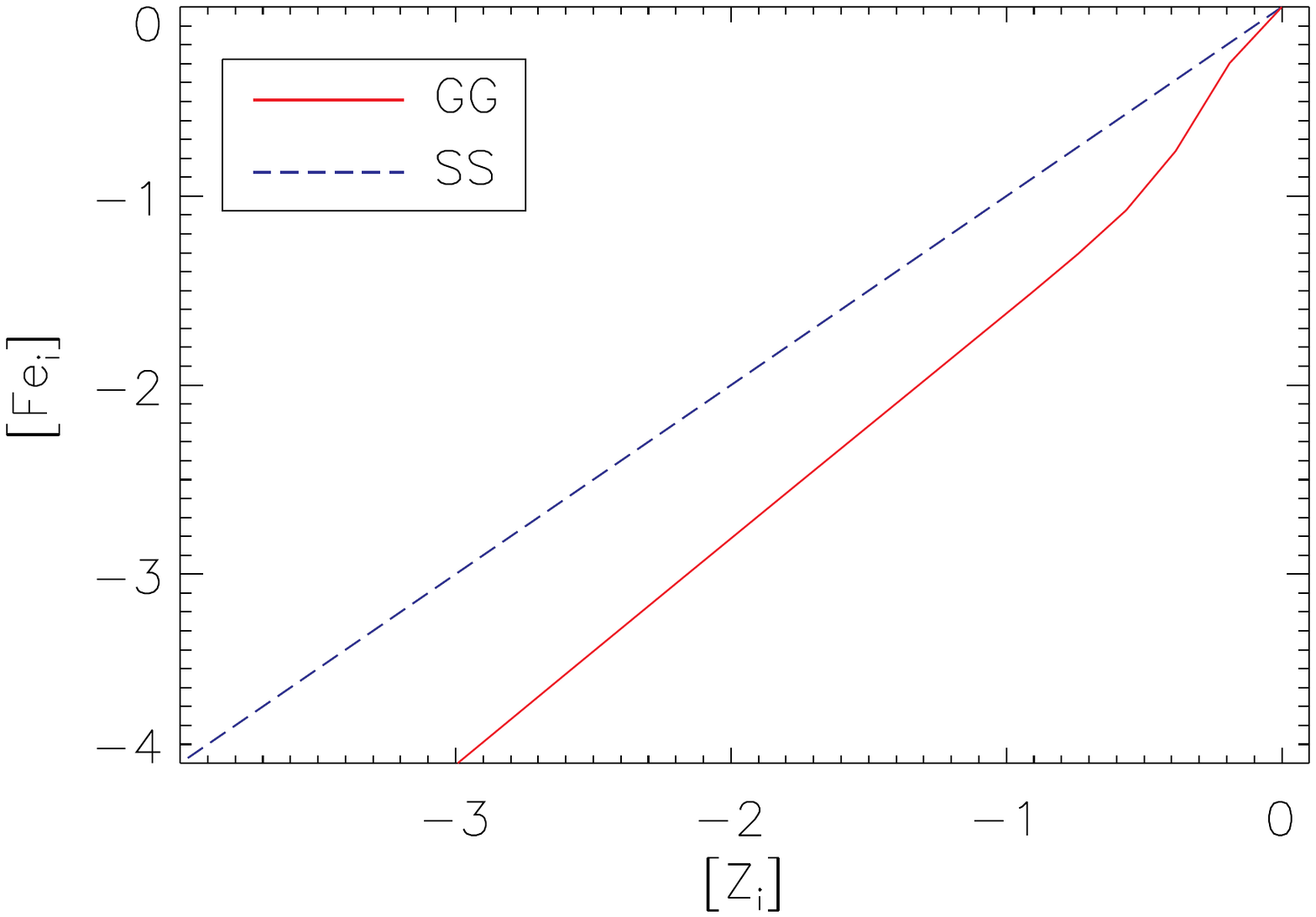}
\hfill
\includegraphics[angle=180,width=0.48\textwidth]{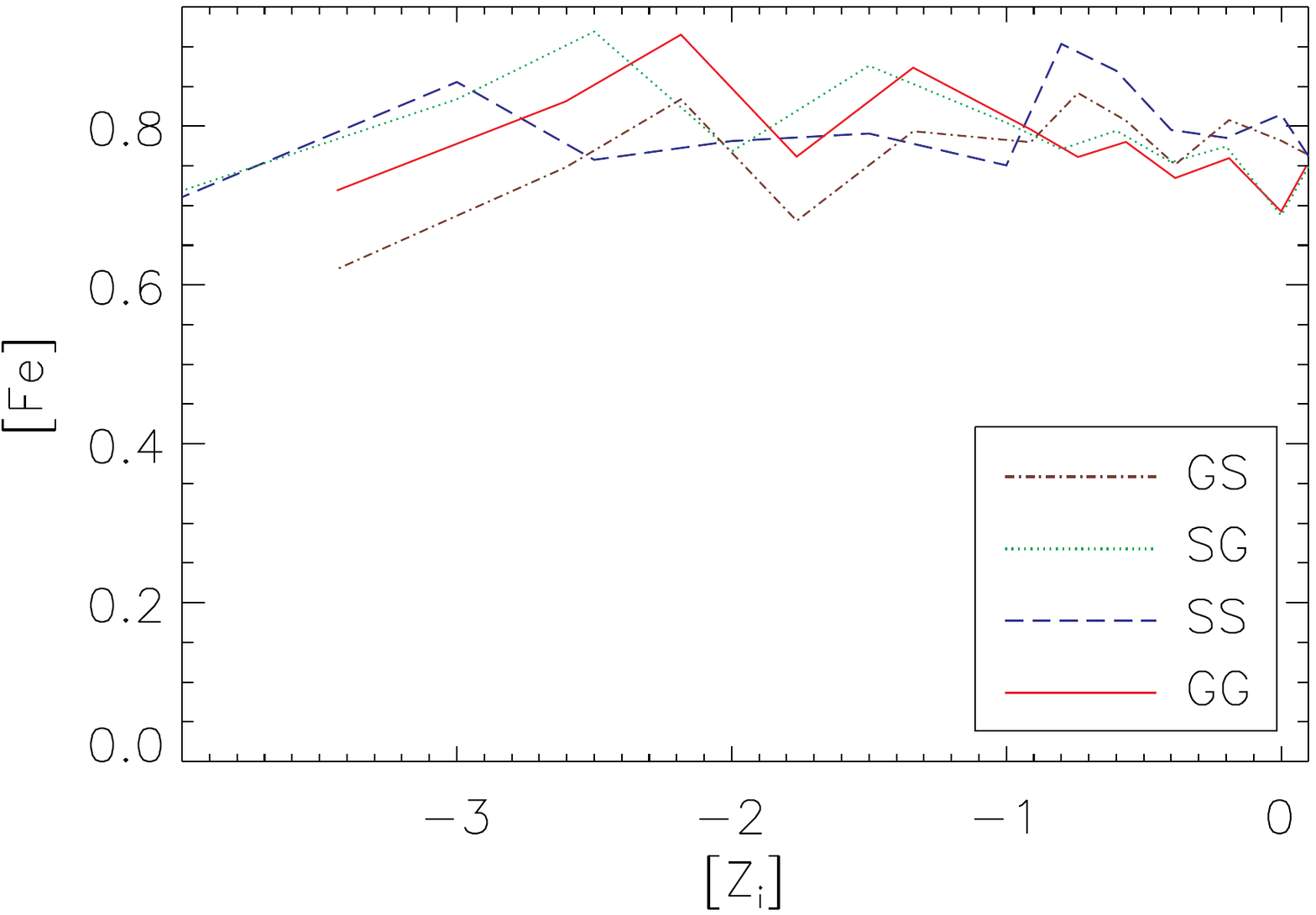}
\caption{\emph{Left:} The initial Fe abundance for both GG (GS) and SS (SG) models, as a function of the initial total metallicity. \emph{Right:} The final Fe yields.\label{fig:Fe}}
\end{figure*}

Due to the treatment of Type Ia SNe in the GG models, the initial Fe abundance is less than the SS models at all sub-solar metallicities (Fig.\,\ref{fig:Fe} \emph{Left}). Note that both GG (SS) and GS (SG) models use the same initial abundances for nucleosynthesis. We find quite similar final Fe yields for all four models (Fig.\,\ref{fig:Fe} \emph{Right}) --- Fe is primary and dominated by explosive yields, hence independent of the initial composition. Since the initial Fe abundances serve as seed isotopes for the weak \emph{s}-process, we expect differences in the weak \emph{s}-only isotope production ($^{70}$Ge, $^{76}$Se, $^{80}$Kr, $^{82}$Kr, $^{86}$Sr, and $^{87}$Sr). Moreover, such differences should not depend solely on the seed abundance, but also on the neutron source. The neutron-producing reaction, $\mathrm{^{14}N\left(\alpha,\gamma\right)^{18}F\left(\beta^{+}\nu_{e}\right)^{18}O\left(\alpha,\gamma\right)^{22}Ne(\alpha,n)^{25}Mg}$, uses $^{14}$N abundances made in the CNO cycle. Whereas CNO burning does not change the overall CNO content, it does re-distribute the abundance between the isotopes involved. Thus a larger initial CNO abundance will result in a corresponding larger $^{14}$N abundance after CNO burning, which becomes the $^{22}$Ne source for neutrons, and can potentially provide more neutrons given sufficient He.

\begin{figure}[th]
\centering
\includegraphics[angle=180,width=\columnwidth]{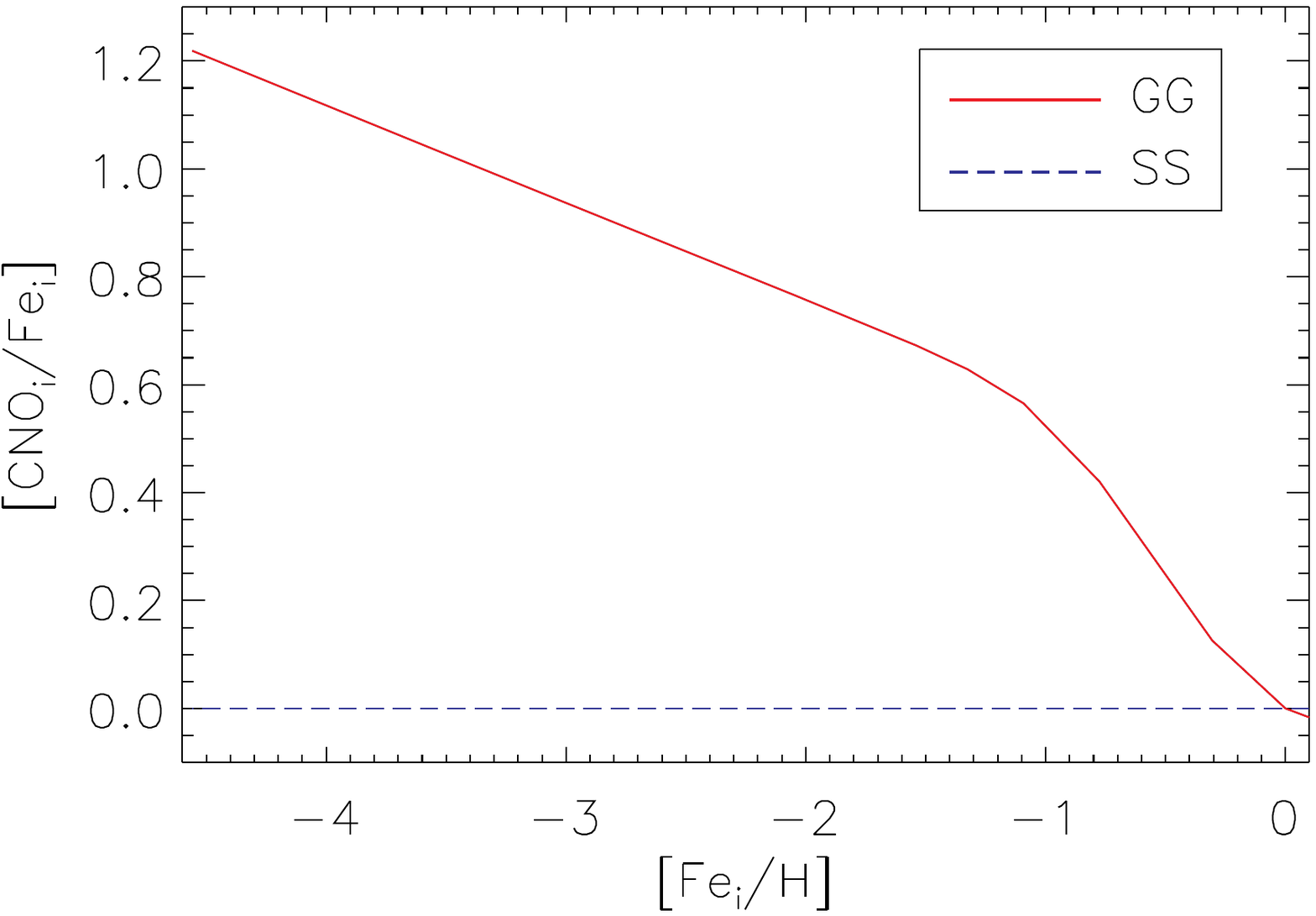}
\figcaption{\label{fig:cno_fe} The initial CNO/Fe ratios as a function of initial
{[}Fe/H{]} for both scaled solar and galactic scaling model compositions.}
\end{figure}

In Fig.\,\ref{fig:cno_fe}, we give the ratio of initial CNO to Fe for both GG and SS models. The GG models have a greater initial CNO/Fe ratio at all sub-solar metallicities, with the largest differences at the lowest metallicities considered. In general, a larger neutron-to-seed ratio can produce more \emph{s}-process abundances, hence one might expect an overproduction of weak \emph{s}-process abundances in the GG models relative to the SS models. In reality, the situation is more complicated. A very large neutron-to-seed ratio can drive the \emph{s}-process abundances to increasingly higher mass numbers. Indeed, this is precisely the description of how low metallicity AGB stars produce the heaviest \emph{s}-only isotopes. Hence, whereas a larger ratio is typically indicative of the \emph{s}-process efficacy, it may not show a constant increased \emph{s}-process abundances for all isotopes.

\subsection{Weak \emph{s}-Process Production\label{ws-process production}}

\begin{figure*}[ht]
\centering
\includegraphics[width=0.8\textwidth,angle=180,origin=c]{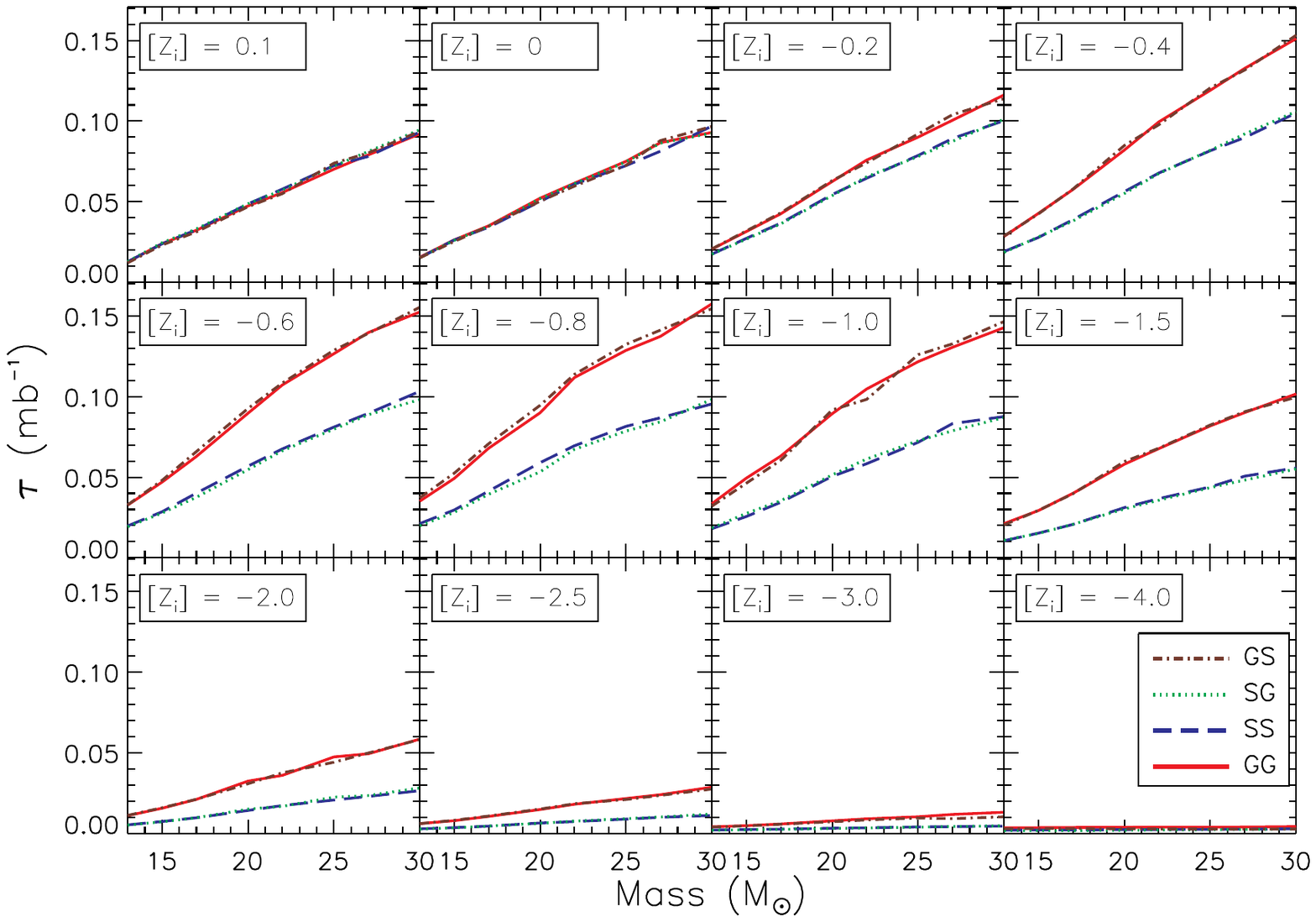}
\caption{\label{fig4-2} The estimated neutron exposures for all models: GG (line), GS (dot-dashed), SS (dashed), and SG (dotted), as a function of stellar mass, at all initial metallicities.}
\end{figure*}

The differences in weak \emph{s}-process production between the GG/GS and SS/SG were found to be due to larger neutron exposures in the core during core-He burning. Moreover, larger mass models experienced a larger neutron exposure relative to lighter mass models. To demonstrate this, we estimated the neutron exposures for all models assuming an \textsuperscript{56}Fe seed. In practice, many isotopes can seed the weak \emph{s}-process, and the \textsuperscript{56}Fe itself can be replenished from neutron captures on \textsuperscript{54}Fe. The choice to use \textsuperscript{56}Fe, as opposed to \textsuperscript{54}Fe, which \emph{cannot} be replenished by neutron capture, is motivated by the large neutron capture cross-section of \textsuperscript{56}Fe and its relatively high abundance. We calculated the neutron exposure using the approximate core \textsuperscript{56}Fe abundance at the beginning and end of core-He burning,

\begin{equation}
\tau\approx-\frac{1}{\sigma}\mathrm{ln\left(\frac{\mathit{X}_{0}^{Fe56}}{\mathit{X}_{f}^{Fe56}}\right)},
\end{equation}

where $\tau$ is the estimated neutron exposure, $\sigma=-11.7\,\mathrm{mb}$ is the Maxwellian-averaged neutron capture cross-section (evaluated at an energy of $30\,\mathrm{keV}$; \citealp{Dillmann2009}), and $X_{\mathrm{0}}^{\mathrm{Fe56}}$ and $X_{\mathrm{f}}^{\mathrm{Fe56}}$ are the initial and final core \textsuperscript{56}Fe abundances, respectively. In Fig.\,\ref{fig4-2}, we give the distribution of neutron exposures for all models at each initial metallicity. At all initial stellar masses, the GG/GS models have larger neutron exposures than the SS/SG. Moreover, the structure makes very little difference, as is expected. This calculation of $\tau$ is only approximate, as convective mixing near the edge of the core makes it difficult to accurately assess the correct Fe abundance to measure (see \S\ref{sec:Mixing} for a discussion).

\begin{figure}[ht]
\centering
\includegraphics[angle=180,width=\columnwidth]{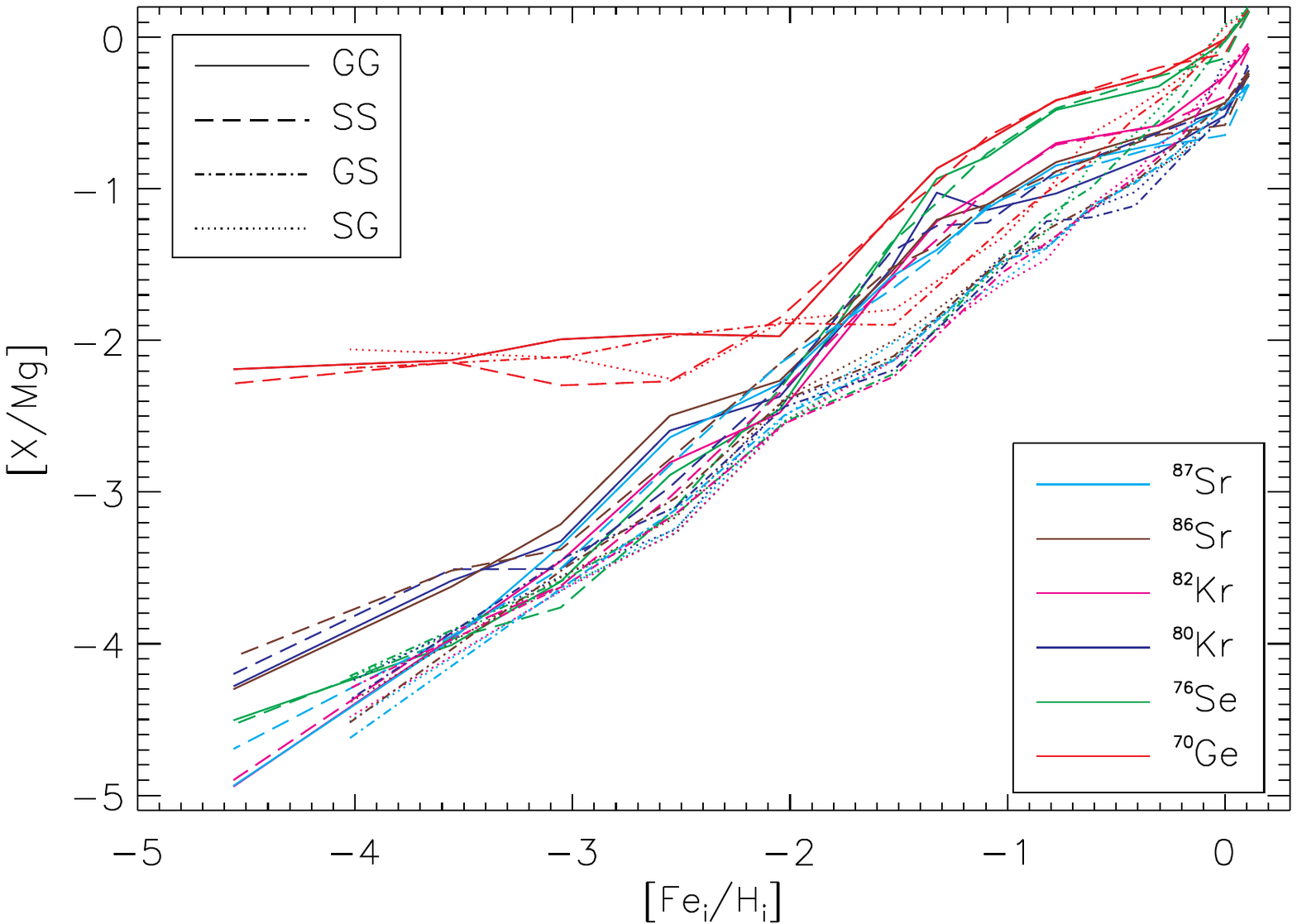}
\figcaption{\label{fig:s_only} The produced weak \emph{s}-only isotope yields as a function of initial [Fe/H] for both SS and GG models. In general, at sub-solar metallicities, the GG models have larger productions than the SS models.}
\end{figure}

In general, we find greater \emph{s}-process production in the GG models at low metallicities ($[Z]\lesssim0$). A comparison is given in Fig.\,\ref{fig:s_only} for the six \emph{s}-only isotopes along the weak \emph{s}-process path. In the higher metallicity ($[Z]\gtrsim0$) models, the \emph{s}-only production becomes greater for SS, at different metallicities for each isotope.

To evaluate which set of models agrees with observations, we computed integrated yields for all models and compared the result to the solar abundances. The integrations were performed over the metallicity interval, and the weak \emph{s}-process yields were linearly interpolated between adjacent models,

\begin{equation}
Y_{i}^{\prime}=\sum_{j}\intop_{z_{j}}^{z_{j+1}}\left(a_{j}\cdot\left[z-z_{j}\right]+Y_{i,j}^{*}\right)\cdot dz,\label{eq:1-1}
\end{equation}

where the IMF interpolated yield mass for isotope \emph{i} at metallicity $z_{j}$ is given by $Y_{i.j}^{*}$, the integrated yield mass for isotope \emph{i} is given by $Y_{i}^{\prime}$, and the interpolated slopes are given by,\emph{ $a_{i,j}=\left(Y_{i,j+1}^{*}-Y_{i,j}^{*}\right)/\left(z_{j+1}-z_{j}\right)$.}The sum runs over the metallicity range, $\left[Z_{i}\right]\in\left(-4,-0.2\right)$. We computed integrated yields for the six weak \emph{s}-only isotopes, as well as the three stable Mg isotopes. Using these results, we find more weak \emph{s}-process production in the GG models, with $\left[\mathrm{s/Mg}\right]_{\mathrm{GG}}=-0.69$, where $\left[\mathrm{s}\right]$ is the logarithm of the sum of the six weak \emph{s}-only isotopes relative to the fraction of their solar values attributed to this process (taken from \citealp{West2013b}). Since the weak \emph{s}-process isotopes have significant contributions from massive star nucleosynthesis, we anticipate more production from model sets with compositions that are indicative of a more accurate chemical history.  Both SG and SS models show less production, with $\left[\mathrm{s/Mg}\right]_{\mathrm{SG}}=-0.92$ and $\left[\mathrm{s/Mg}\right]_{\mathrm{SS}}=-0.90$, and the GS models are in between, with $\left[\mathrm{s/Mg}\right]_{\mathrm{GS}}=-0.77$. The composition used for computing the nucleosynthesis appears more important than the composition used for the structure, with the GCH models favored over scaled-solar when compared to observations. Caution should be taken in interpreting the above preliminary analysis, however, since a proper chemical evolution computation was not performed for it.

\subsection{OMEGA Computations}

\begin{figure*}[htp]
\centering
\subfloat{%
  \includegraphics[clip,width=0.8\textwidth]{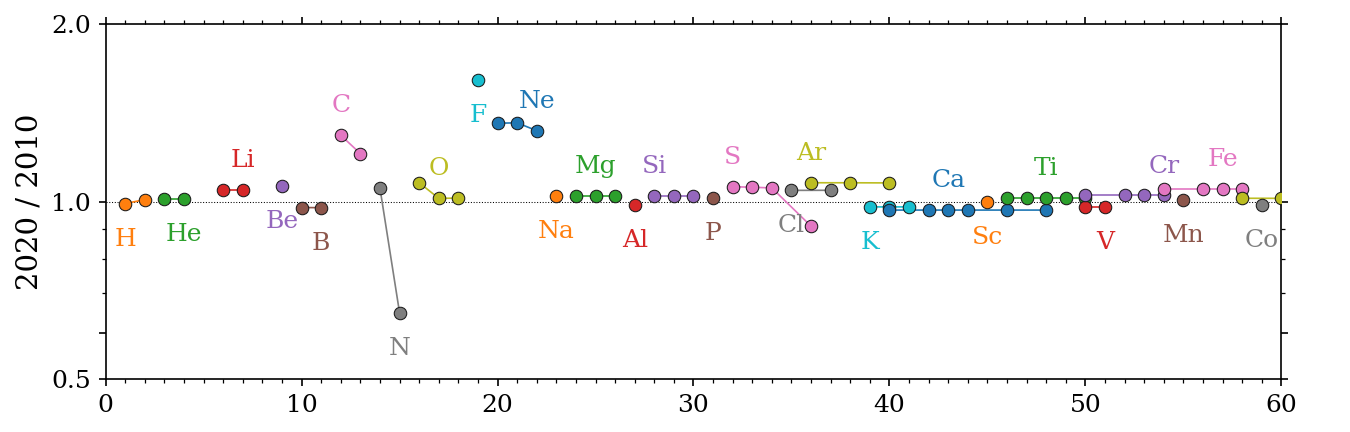}%
}

\subfloat{%
  \includegraphics[clip,width=0.8\textwidth]{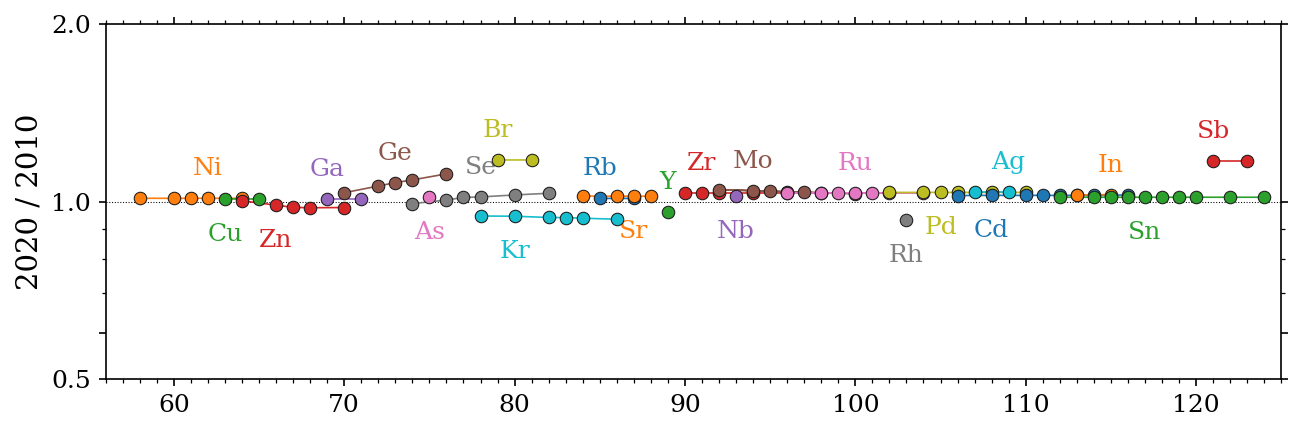}%
}

\subfloat{%
  \includegraphics[clip,width=0.8\textwidth]{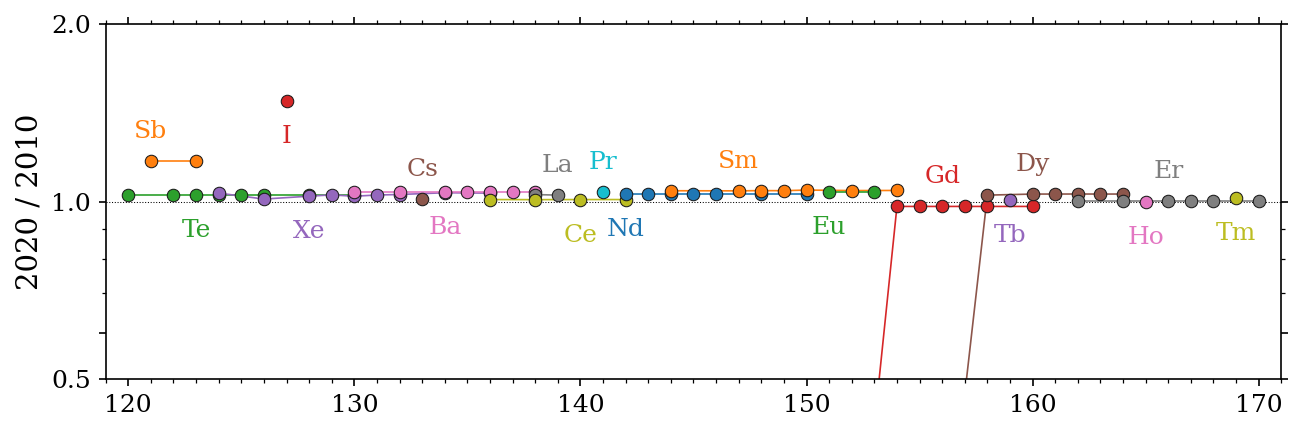}%
}

\subfloat{%
  \includegraphics[clip,width=0.8\textwidth]{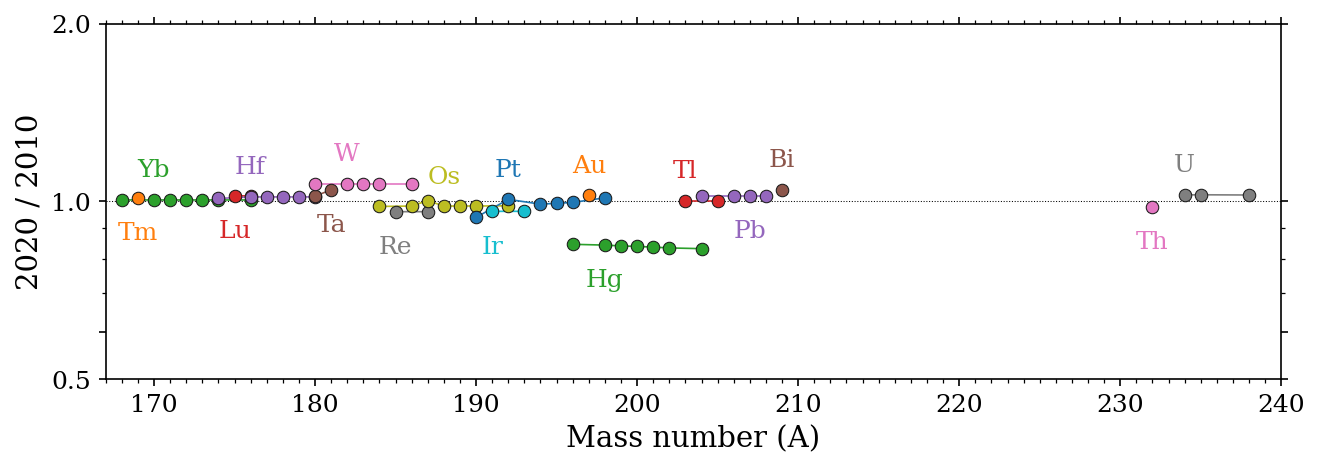}%
}
\caption{\label{fig:lod_ratio} The ratio of 2020/2010 solar abundances provided by \citet{Lodders2020,Lodders2010}. The isotopic abundance ratios not shown and their values are $^{152}\textrm{Gd}\approx0.177$ and $^{156}\textrm{Dy}\approx0.217$}
\end{figure*}

Directly comparing integrated yields from the GG, SS, SG, and GS models to solar observations provides limited results. In principle, these yields should be the ingredients for a galactic chemical evolution code, and similar efforts have been previously made (e.g., \citealp{Cote2016}). We used the OMEGA code to compute the chemical evolution for all four sets. The solar abundances used hereafter come from an updated set provided by \citet{Lodders2020}. The isotopic ratios between the \citet{Lodders2020} and \citet{Lodders2010} abundances are shown in Fig.\,\ref{fig:lod_ratio}. The largest deviations from the 2010 abundances include:  $^{12,13}\textrm{C}$, $^{15}\textrm{N}$, $^{19}\textrm{F}$, $^{20,21,22}\textrm{Ne}$, and $^{127}\textrm{I}$. Both $^{152}\textrm{Gd}$ and $^{156}\textrm{Dy}$ are $p$-isotopes, and so small updates to their abundances by number (normalized to $10^6$ silicon atoms) can result in large ratio differences.

The OMEGA code is a one-zone model for the chemical evolution of galaxies (for description and discussion, see, \citealp{Cote2017,Cote2018}). The computed elemental evolutions of selected elements for the SS and GG models are given in the Appendix (see Figs.\,\ref{fig:GCE1} \& \ref{fig:GCE2}). All show similar trends, which suggests an insensitivity of the GCE model against the different isotopic inputs. At low metallicities, massive star evolution dominates metal production. The explosive environments that end these stars re-process both some fraction of the initial composition and also abundances produced during hydrostatic burning, both of which are impacted by different initial composition sets. Such explosive processing may partially erase differences in isotopic distributions produced in previous stellar generations to some degree. Of course, some fraction of the initial composition also returns to the ISM unprocessed, although the particular ratios are difficult to disentangle. As a consequence, it may be that re-processing can reduce differing isotopic distributions made during hydrostatic burning into similar compositions dominated by alpha isotopes, effectively normalizing final compositions across multiple chemical generations.

\begin{figure*}[h]
\centering
\includegraphics[width=0.8\textwidth,origin=c]{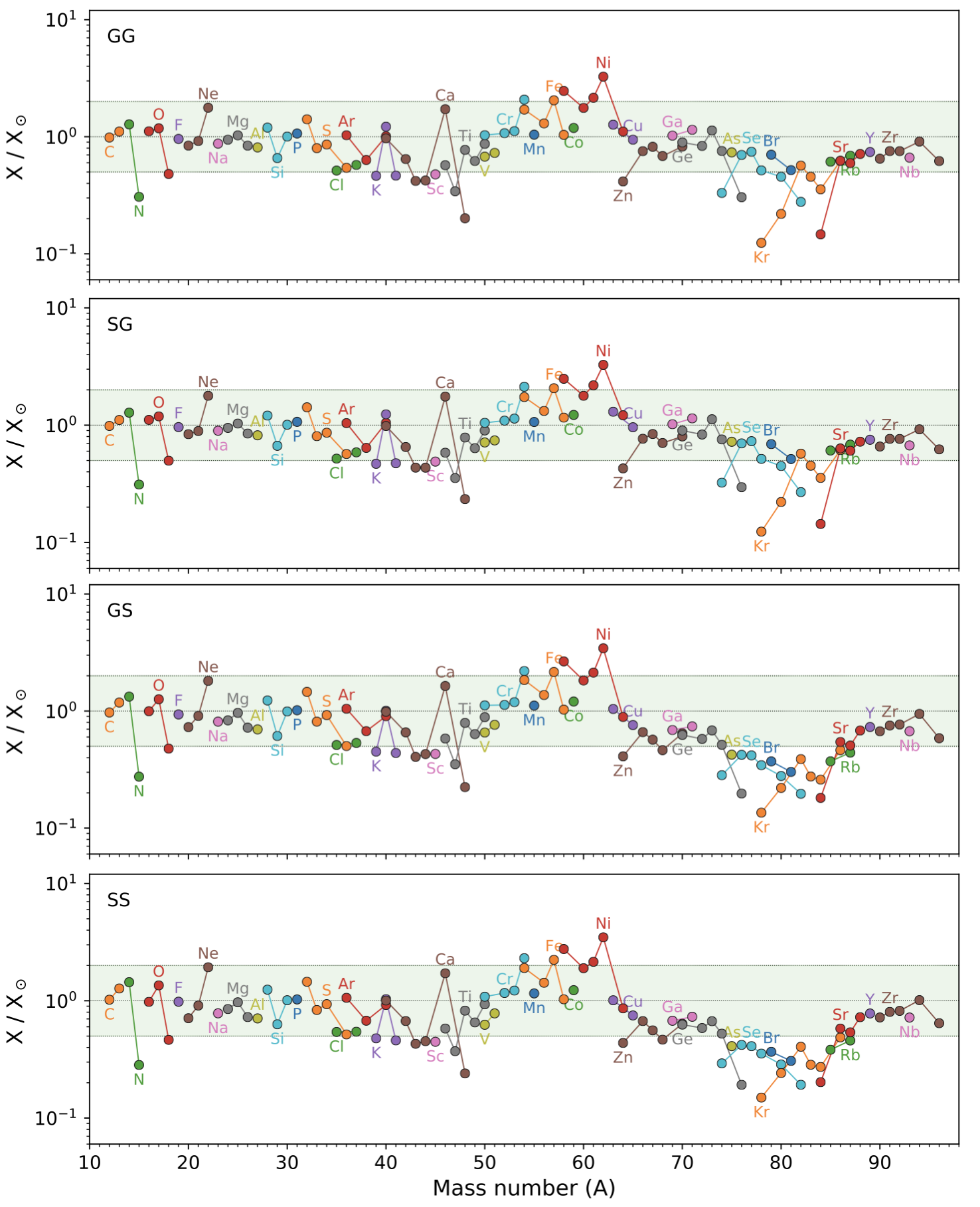}
\figcaption{\label{fig:gce_all} The final abundances for GG SS, SG and GS model sets relative to the \citet{Lodders2020} solar abundances.}
\end{figure*}

The final abundances for all four sets are shown against solar in Fig.\,\ref{fig:gce_all}. The apparent difference between model sets is larger production of isotopes in the approximate range $70 \le A \le 85$ in the GG and SG compared to SS and GS. Evidently, differences in the isotopic composition of the stellar structure correlates to larger production in this range more than the composition used in computing the nucleosynthesis.

Notable is the overproduction of $^{62}$Ni, which has been observed previously (see, for e.g., \citealp{Rauscher2002}). This may be an artifact of uncertain neutron capture cross-sections used in the reaction $^{62}Ni(n,\gamma)^{63}Ni$, which has been shown to impact not only the $^{62}$Ni abundance, but also the resulting weak \emph{s}-process chain that follows \citep{Nassar2005}.

\begin{figure*}[ht]
\centering
\includegraphics[width=0.8\textwidth,origin=c]{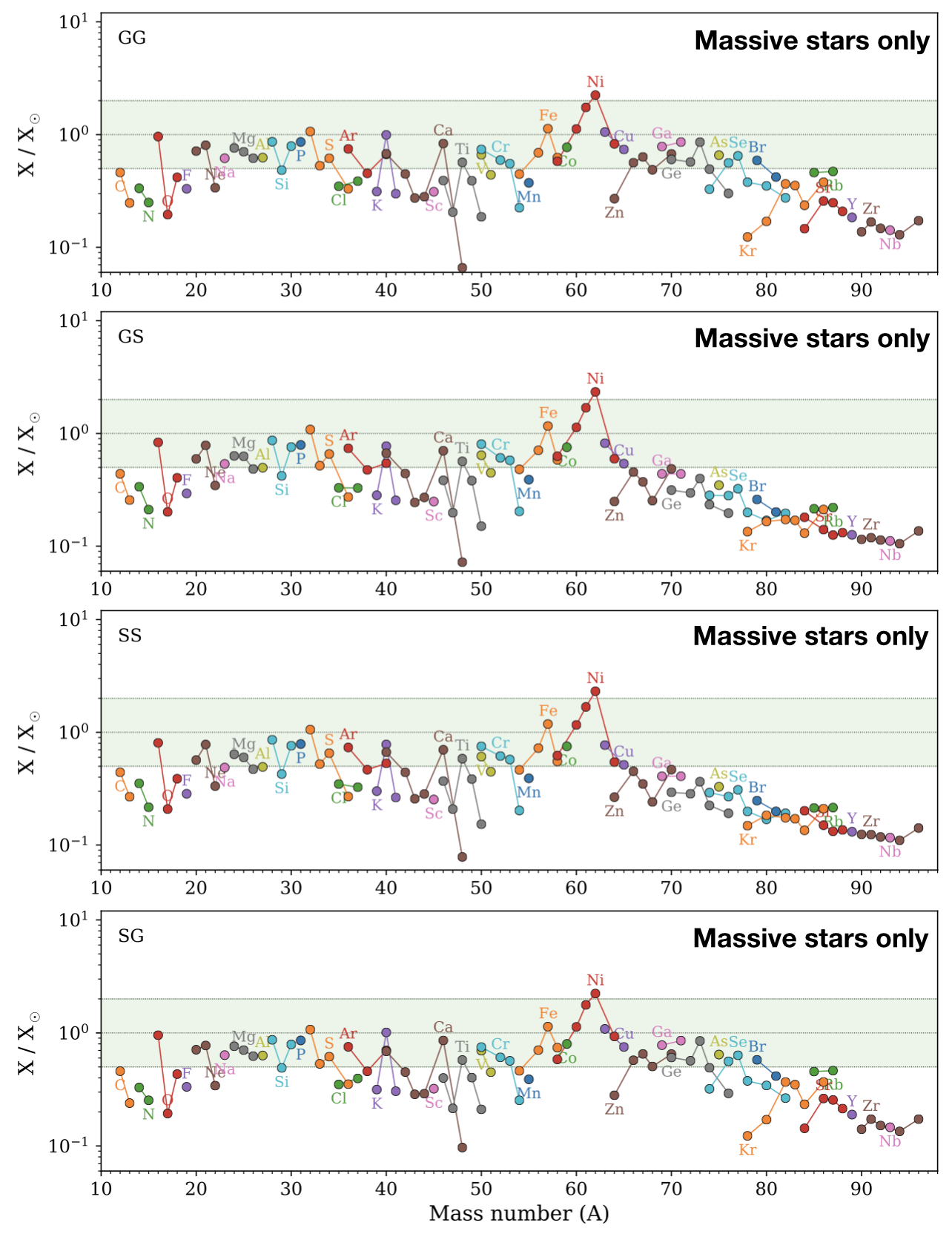}
\figcaption{\label{fig:gce_massive} The final abundances for GG SS, SG and GS model sets relative to the \citet{Lodders2020} solar abundances. Only contributions from massive star nucleosynthesis are shown.}
\end{figure*}

\begin{figure*}[ht]
\centering
\includegraphics[width=0.8\textwidth,origin=c]{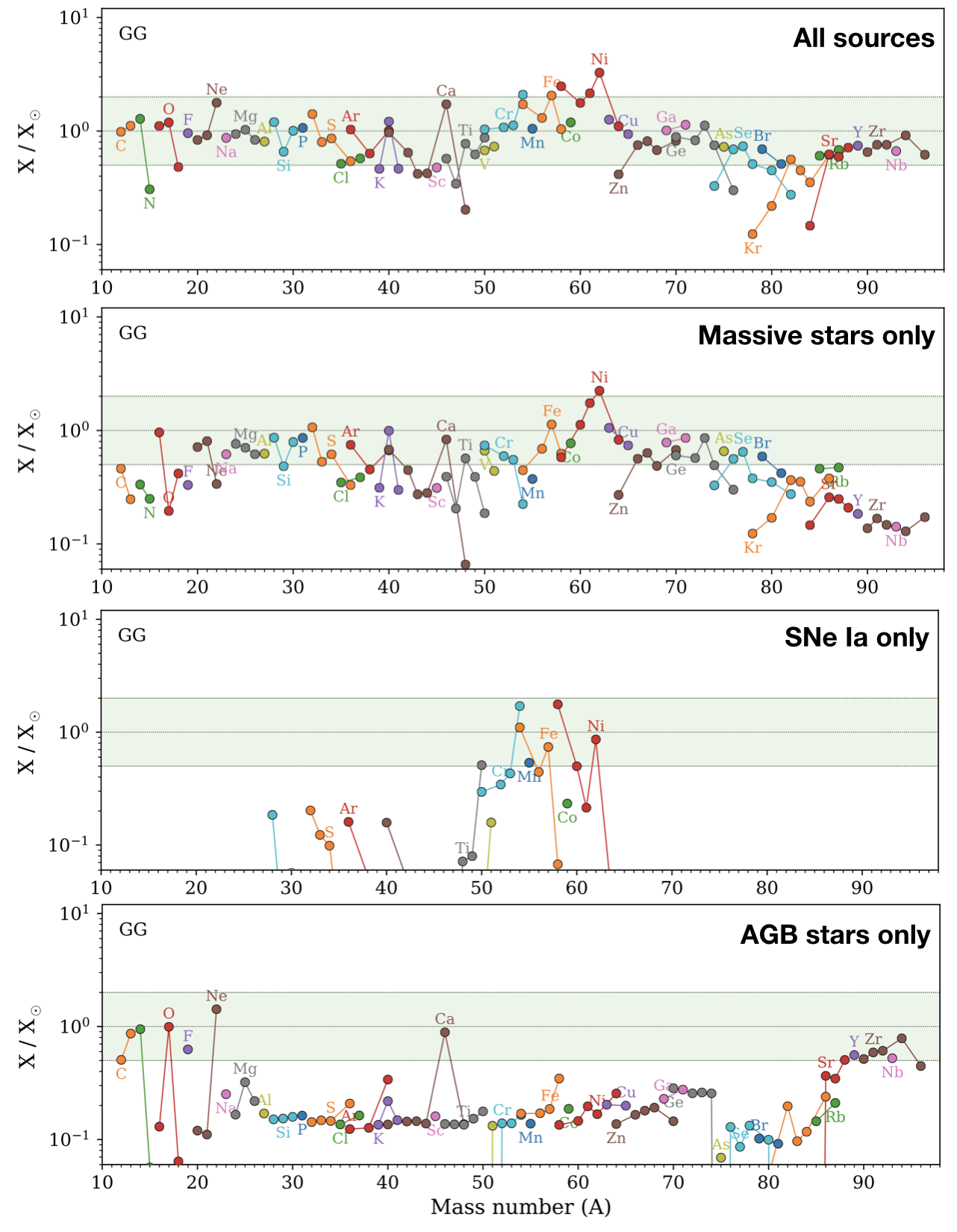}
\figcaption{\label{fig:gce_GG} The final abundances for GG model sets parsed into AGB, Type Ia, and massive star contributions, relative to the \citet{Lodders2020} solar abundances.}
\end{figure*}

Despite this, we still find an increased production of isotopes in the weak \emph{s}-process range to result from massive star nucleosynthesis. A clearer delineation of this production is found once the AGB and Type Ia contributions are removed, see Fig.\,\ref{fig:gce_massive}. A comparison between AGB, Type Ia, and massive stars is given for GG in Fig.\,\ref{fig:gce_GG}.

As shown in Fig.\,\ref{fig:gce_massive}, increased weak \emph{s}-process production is found in the range $70 \le A \le 85$ for the SG \& GG models (which use GCH for computing the stellar structure), compared to less production for the GS \& SS models (which use scaled-solar for computing the stellar structure). 

As shown in Fig.\,\ref{fig:gce_GG}, we can identify expected nucleosynthesis trends from massive, AGB, and Type Ia. The AGB abundances show the main \emph{s}-process productions beginning at $A\gtrsim 85$ with Sr production from AGB being greater than massive by about $0.2\,$dex.

\begin{figure*}[ht]
\centering
\includegraphics[width=0.8\textwidth,origin=c]{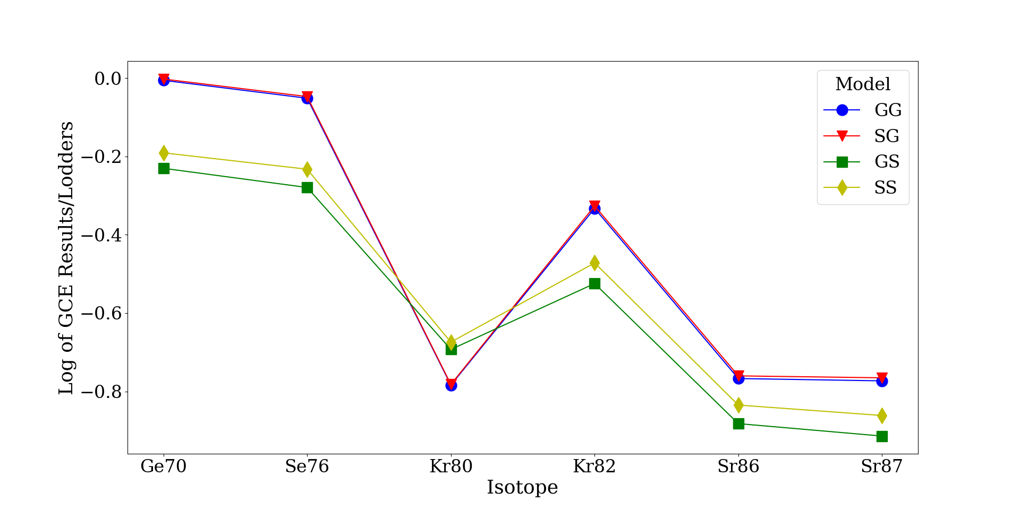}
\figcaption{\label{fig:sonly_4} The ratio of final GCE abundances to the \cite{Lodders2020} solar abundances for the six \emph{s}-only isotopes along the weak \emph{s}-process path. Note: Values were multiplied by \citet{Lodders2010} solar decomposition fractions for the weak s-process for each isotope, as done in \citet{West2013b}.}
\end{figure*}

To better discuss the impact on weak \emph{s}-process production, the abundance ratios from the four computed GCE models are shown for the six weak \emph{s}-only isotopes, see Fig.\,\ref{fig:sonly_4}. Note that all four GCE model set results remained within an order of magnitude of the \cite{Lodders2020} data, and under-produced the weak \emph{s}-only isotopes except $^{70}$Ge, which was consistent with its solar value for the GG and SG models. The GCE results for each combination of the scaled-solar and GCH models showed greater differences resulting from stellar structure than from nucleosynthesis in the six weak \emph{s}-process only isotopes. As shown in Fig.\,\ref{fig:sonly_4}, the final abundances for the six weak \emph{s}-only isotopes track according to the structure, GG with SG, and GS with SS. Moreover, there was less variation in the results for the GG and SG models (which shared the GCH model as an input for stellar structure) compared to the GS and SS models (which shared the scaled-solar model as an input for stellar structure). The GG and SG results were overall closer to solar abundances of the weak \emph{s}-only isotopes, with the exception of $^{80}$Kr, which was under-produced in the GG and SG results by a greater margin than in the GS and SS results. 

\begin{figure*}[ht]
\centering
\includegraphics[width=0.8\textwidth,origin=c]{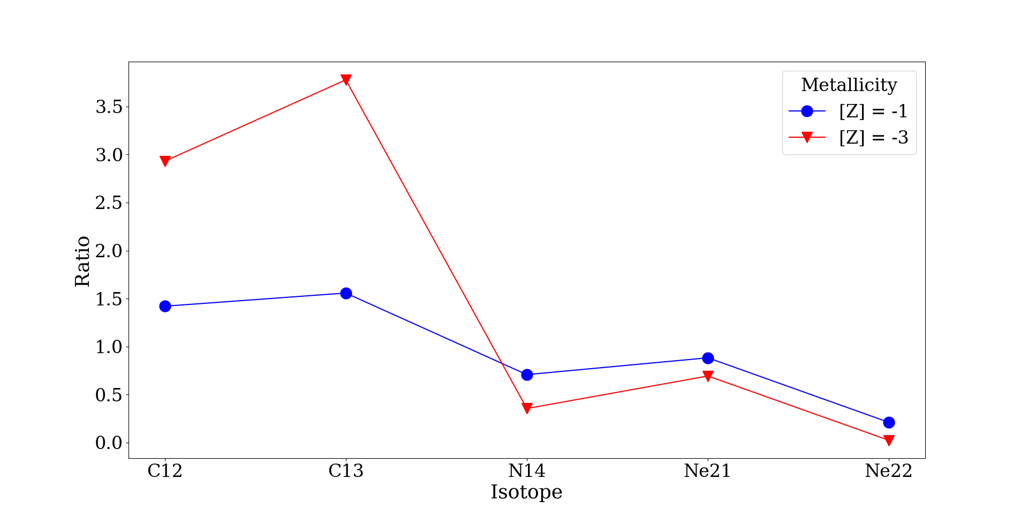}
\figcaption{\label{fig:neutron_sources_2} The ratio of GCH to scaled-solar initial abundances for relevant neutron sources at $[Z]=-1$ and $[Z]=-3$. These inital abundances from the scaled-solar and GCH models were used as inputs in the stellar simulations.}
\end{figure*}

Further discussion of the production of the weak-\emph{s} process across the GCE results is addressed through an analysis of neutron sources, seeds, and poisons. Initial neutron source abundance ratios were found at $[Z]=-1$ and $[Z]=-3$, see Fig.\,\ref{fig:neutron_sources_2}. The selected neutron source isotopes are $^{12}$C, $^{13}$C, $^{14}$N, $^{21}$Ne, and $^{22}$Ne. The $^{18}$O isotope was not included in this set because its initial abundance was not expected to substantially impact the structure of the stellar models.

\begin{figure*}[ht]
\centering
\includegraphics[width=0.8\textwidth,origin=c]{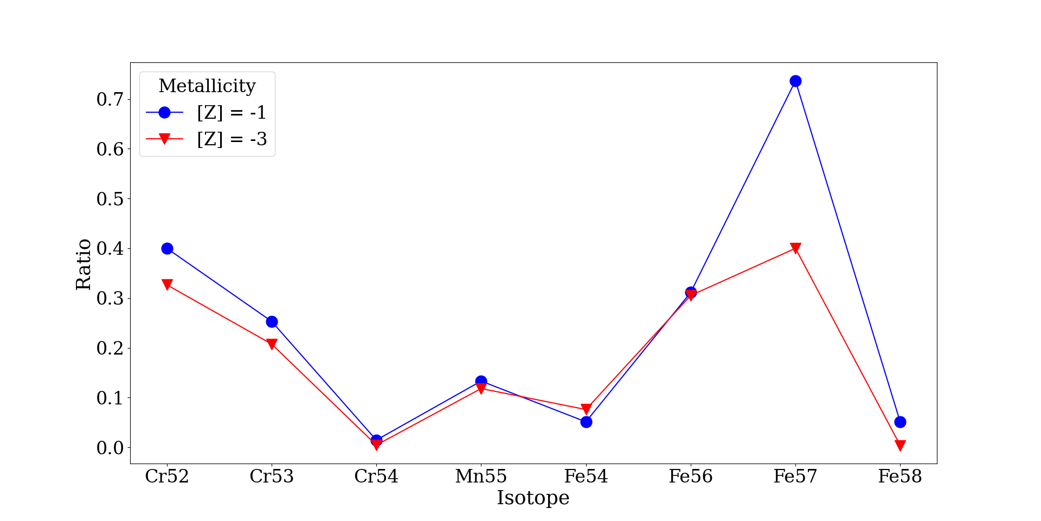}
\figcaption{\label{fig:neutron_seed} The ratio of GCH to scaled solar initial abundances for relevant neutron capture seed nuclei at $[Z]=-1$ and $[Z]=-3$. These inital abundances from the scaled solar and GCH models were used as inputs in the stellar simulations.}
\end{figure*}

The ratio of input abundances for neutron capture seed nuclei between GCH and scaled-solar models at $[Z]=-1$ and $[Z]=-3$ can be found in Fig.\,\ref{fig:neutron_seed}. The selected neutron seeds are eight stable iron-peak nuclei, hence have relatively large abundances and require fewer neutrons captured to move beyond iron into weak \emph{s}-process products. The seed nuclei considered are $^{52}$Cr, $^{53}$Cr, $^{54}$Cr, $^{55}$Mn, $^{54}$Fe, $^{56}$Fe, $^{57}$Fe, and $^{58}$Fe.

\begin{figure*}[ht]
\centering
\includegraphics[width=0.8\textwidth,origin=c]{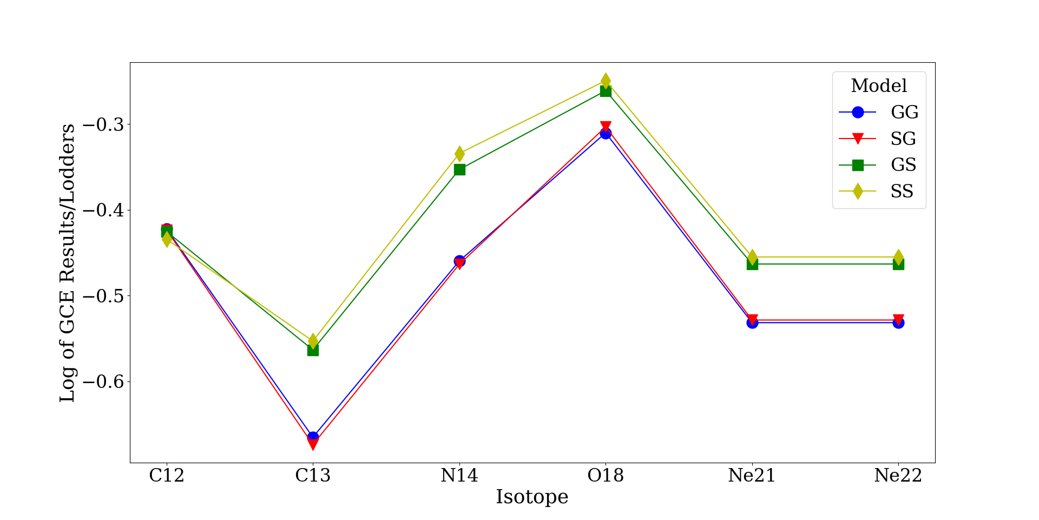}
\figcaption{\label{fig:neutron_sources_4} The final GCE model abundances for relevant neutron sources for each set of initial abundances, relative to the \citet{Lodders2020} solar abundances.}
\end{figure*}

The selected neutron sources identified are $^{12}$C, $^{13}$C, $^{14}$N, $^{18}$O, $^{21}$Ne, and $^{22}$Ne. The $^{13}$C and $^{22}$Ne abundances drive the two primary neutron producing reactions for the weak \emph{s}-process. The $^{14}$N and $^{21}$Ne isotopes are involved in neutron producing reactions, and $^{18}$O is involved in $^{22}$Ne production (e.g., \citealp{Pignatari2010}). Note that, while $^{12}$C is not directly involved in important neutron producing reactions, it is a key component in $^{13}$C production and can determine how much of a star’s carbon shell is convective (see, for e.g., \citealp{West2013a}). See Fig.\,\ref{fig:neutron_sources_4} for the neutron source abundances from the GG/GS/SG/SS GCE models. As shown in Fig.\,\ref{fig:neutron_sources_4}, all four sets of results under-produced neutron sources. The GG and SG results had lower neutron source abundances compared to the GS and SS results, with the SS results having the greatest neutron source abundance.

\begin{figure*}[ht]
\centering
\includegraphics[width=0.8\textwidth,origin=c]{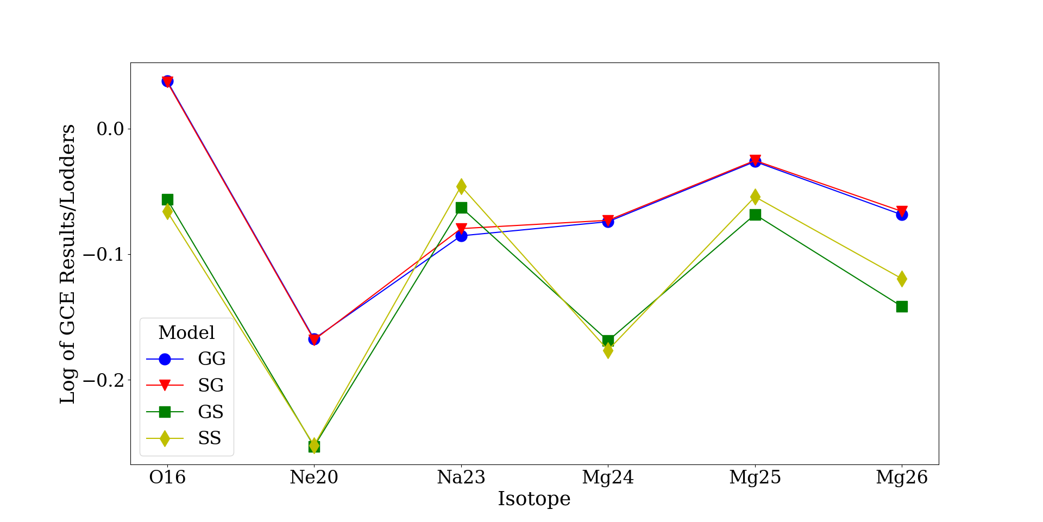}
\figcaption{\label{fig:neutron_poisons_4} The final GCE model abundances for relevant neutron poisons for each set of initial abundances, relative to the \citet{Lodders2020} solar abundances.}
\end{figure*}

The selected neutron poisons identified are $^{16}$O, $^{20}$Ne, $^{23}$Na, $^{24}$Mg, $^{25}$Mg, and $^{26}$Mg. High concentrations of $^{16}$O can act as a strong neutron poison at low metallicities, and $^{20}$Ne can, in some circumstances, act as a light neutron poison. $^{23}$Na can be an important neutron poison in the carbon shell, and $^{24}$Mg, $^{25}$Mg, and $^{26}$Mg all take part in neutron capture reactions during the same burning stages as the weak \emph{s}-process, with $^{25}$Mg in particular having a large neutron capture cross-section (see, for e.g., \citealp{Pignatari2010,Rayet2000}). See Fig.\,\ref{fig:neutron_poisons_4} for the neutron poison abundances from all GCE models.

All four models under-produced the neutron poisons considered (though still within an order of magnitude of the \citealp{Lodders2020} abundances), with the exception of $^{16}$O for the GG/SG models (see Fig.\,\ref{fig:neutron_poisons_4}). The GG and SG results produced more of every neutron poison except for $^{23}$Na, though the abundances are rather similar across the models for this isotope. The GG/SG and GS/SS results again track each other, again indicating that the abundances used for determining the structure are a more important factor for driving the resulting nucleosynthesis, with slightly more pronounced differences between the GS and SS results than the GG and SG. Aside from $^{12}$C and $^{20}$Ne, all neutron poisons were produced in greater ratios than neutron sources in the final GCE model yields.

The initial abundances for the GCH model show notable differences from the scaled-solar model in neutron sources and seed nuclei. At $[Z]=-1$, isotopes involved in neutron producing reactions were produced up to $\approx 1.5$ times more in the GCH model than the scaled-solar model for the $^{12}$C and $^{13}$C isotopes, whereas the other seeds were produced less in the GCH model (see Fig.\,\ref{fig:neutron_sources_2}). At $[Z]=-3$, the $^{12}$C and $^{13}$C isotopes were produced $\approx 3$ and $\approx 4$ times more in the GCH model than the scaled solar model, respectively. As shown in Fig.\,\ref{fig:neutron_seed}, the GCH model produced less seed nuclei than the scaled solar model at both $[Z]=-1$ and $[Z]=-3$, producing up to two orders of magnitude less of some isotopes than the scaled solar model. The only substantial difference in seed nuclei abundance ratios between the two metallicities was a nearly twofold decrease in the abundance of $^{57}$Fe going from $[Z]=-1$ to $[Z]=-3$.

The increased weak \emph{s}-only isotope production in the models with GCH abundances for the structure possibly resulted from increased initial abundances of carbon in the stellar models. This can lead to a larger convective carbon shell, and hence greater weak \emph{s}-process occurring there (see, for e.g., \citealp{Pignatari2010,West2013a}). This increased initial carbon abundance is greater at low metallicities, where the neutron-to-seed ratio is higher. As a result, this may seem as though this could drive weak-\emph{s} production to higher mass numbers, though for a different reason than suggested by, for e.g., \cite{Frost2022}. Nevertheless, this is not observed in the final abundances shown in Fig.\,\ref{fig:gce_massive}.

\subsection{Machine Learning Parameter Fitting Process} 

The fitting performed in \citet{West2013b} relied first on $[\mathrm{Mg}/\mathrm{Fe}]$ observational data to fit the $f$, $a$, and $b$ parameters, and then used these results to fit the other free parameters in a stochastic process. In this sense, the fitting did not handle all data equally, and although it leveraged the most plentiful data first, it is possible that different fit parameters might result if the fitting was done in an alternate order.

To atone for this possible bias, a grid search algorithm was performed which systematically explores a predefined range of parameter values, adjusting and evaluating each combination based on performance metrics. This technique was applied to the entirety of the data from \cite{West2013b}, which comprised $14\mathord,249$ astrophysical data points. These points encompass the Magnesium and Europium data, which contributed $2\mathord,165$ data points used as the test set, and an additional $12\mathord,084$ data points taken from $48$ other elements as a training set in the method. These data were collected from \cite{Frebel2010,Soubiran2005,Taylor2003}, and data points were removed from binary systems; the data set used was matched with \cite{West2013b} to permit a robust comparison of fit parameter results.

The resulting best-fit parameters from the grid search are compared to the values previously found by \cite{West2013b} in Table\,\ref{tab:params}.

\begin{table}[!htb]\label{table1}
    \centering
    \begin{tabular}{ c c } 
 Parameter & Best-fit Value \\ 
 \hline
$a$           &4.93(5.024)         \\
$b$           &2.82(2.722)          \\
$f$           &0.69(0.693)          \\
$p$           &0.84(0.938)          \\
$h$           &1.41(1.509)          \\
$l$           &1.13(1.227)         \\
$w$           &1.13(1.230)          \\

\end{tabular}
    \caption[Found Parameter Values]{Final results after machine learning algorithms with best-fit values and comparisons with values from previous work (shown in brackets).}
    \label{tab:params}
\end{table}

The updated best-fit values found were largely consistent, with the largest percent difference found for the parameters $p$ of $11\%$. The parameters for the ``strong'' \emph{s}-scaling function (see \citealp{West2013b}) was not included in the re-fitting, as it only has four data points constraining the parameter values. The full GCH model with updated parameter values is now available online at \url{https://isotest-954ca45d7c8d.herokuapp.com/}. Machine-readable data can be downloaded for each isotope separated by process across a range of metallicities, for use in future nucleosynthesis studies. 

\section{Conclusion}
\label{sec:Conclusion}

We performed a preliminary analysis for a grid of stellar models across the initial metallicity range $-4.0\leq[Z_{i}]\leq+0.1$ for 8 different initial masses, using inputs from our scaling (GCH) model and linearly interpolated scaled-solar abundances. Four different sets of these models were run, in order to vary the inputs used for computing the structure and the nucleosynthesis across the two possibilities.

The first results show an increased production for the weak \emph{s}-process abundances in the GCH models, in better agreement to solar observations. Next, GCE models were completed to assess whether the the initial abundances for computing the structure of the stellar models or computing the nucleosynthesis of the stellar models made a bigger impact on the GCE results when compared to solar observations. As shown in Fig.\,\ref{fig:gce_all}, the salient difference between model sets is larger production of isotopes in the approximate range $70 \le A \le 85$ in the GG and SG compared to SS and GS. Hence, the larger production results from GCH abundances in the stellar structure, irrespective of which initial abundance set is used in computing the nucleosynthesis. This trend was also found among the final abundances for the neutron sources and poisons.

The increased weak \emph{s}-only isotope production found in the GG/SG models possibly resulted from increased initial carbon abundances, which promotes \emph{s}-process production by hosting a larger convective carbon-burning shell. Hence, the GCH model providing results that better agree with solar observations suggests massive star simulations have previously been under-producing \emph{s}-process products, perhaps especially at low metallicities. Moreover, the impact on final results from the stellar structure does not suggest massive stars can synthesize \emph{s}-products beyond the previously believed approximate threshold of $A\approx 100$. 

Finally, a machine learning method was used to verify the fit parameters used to determine the initial abundances used in this work that were adopted from \citet{West2013b}. The updated best-fit values found were largely consistent, with the largest notable difference ($11\%$) for the primary process scaling parameter. The full results and updated model are now available online at \url{https://isotest-954ca45d7c8d.herokuapp.com/}. Data can be downloaded by process across a range of metallicities for all $287$ stable isotopes.

One concern persists despite leveraging machine learning, which is the (standardized) use of fitting the model to logarithmic data. Whereas it makes sense to depict astrophysical abundances on logarithmic axes, performing fits in this space handles abundances separated by orders of magnitude with comparable weights in the fitting process. Ideally, one would complete the fits using ``absolute" (linear) abundances. This should be done and at least offered as an option. 

As a final remark, we note a possible sample bias in abundance data where individual stars with peculiar abundances might skew the resulting averages too high or low. One option could be to use metallicity-complete samples, but this then likely reduces the sample size, thus risking the resolution of a robust average. 

\acknowledgements

AH has been supported, in part, by the Australian Research Council through a Future Fellowship (FT120100363) and by a grant from Science and Technology Commission of Shanghai Municipality (Grants No.16DZ2260200) and National Natural Science Foundation of China (Grants No.11655002). BC acknowledges support by the ERC Consolidator Grant (Hungary) funding scheme (Project RADIOSTAR, G.A. n. 724560), and by the National Science Foundation (NSF, USA) under grant No. PHY-1430152 (JINA Center for the Evolution of the Elements)

\appendix

\section{Mixing}
\label{sec:Mixing}

\begin{figure}[ht]
\centering
\includegraphics[angle=180,width=0.5\textwidth]{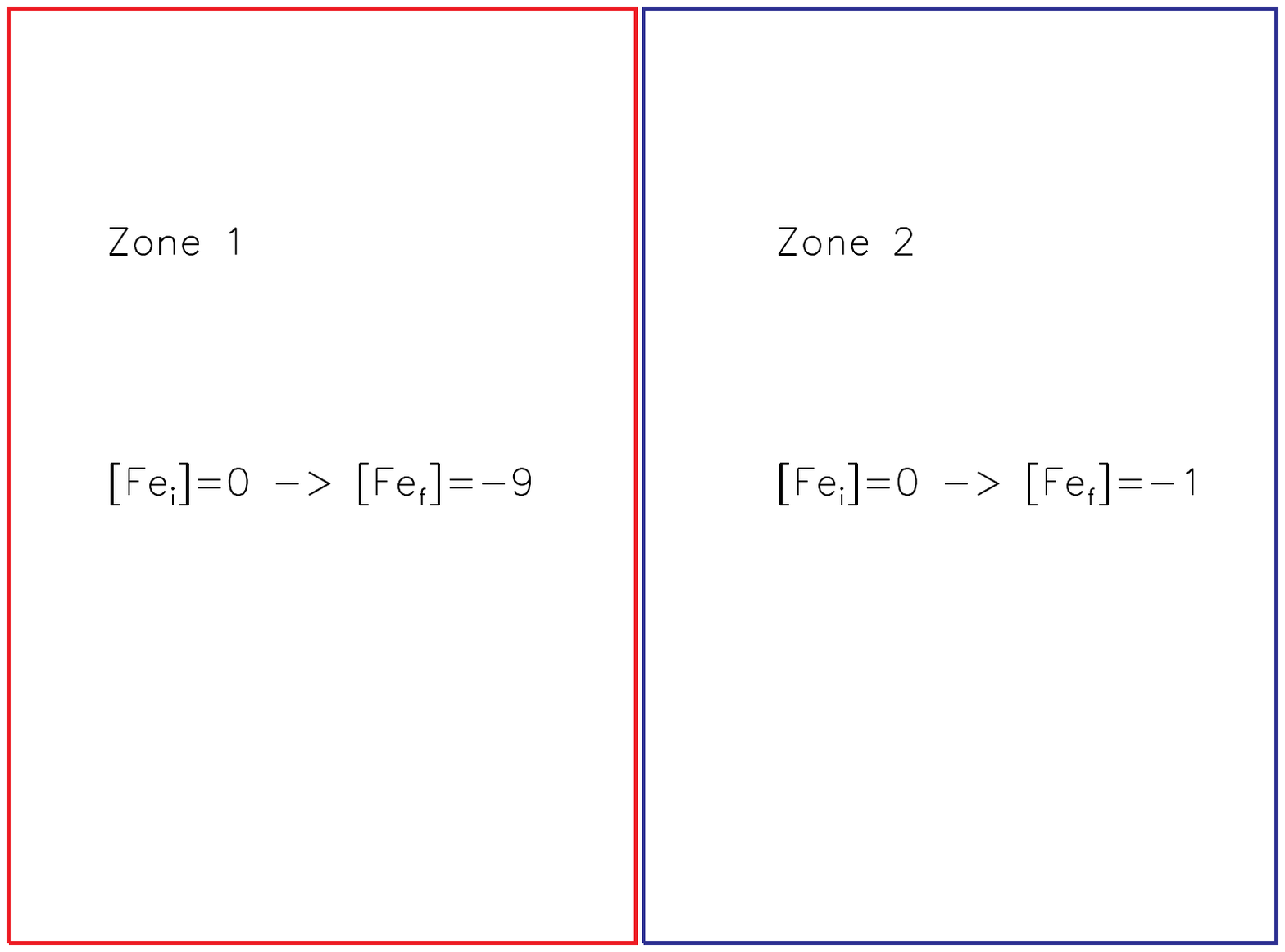}
\caption{\label{fig:zones} Two neighbouring zones of equal mass, experiencing different neutron exposures.}
\end{figure}

The calculations in Section\,\ref{ws-process production} are only approximate, as mixing at the edge of the core can hide the neutron exposures. To illustrate this, consider two neighbouring compositions, Zone 1 and Zone 2, of equal mass, $M$, that begin with same initial abundance of Fe. Suppose Zone 1 experiences quite drastic neutron exposures,  whereas Zone 2 experiences substantially less (Fig.\,\ref{fig:zones}). If these zones later convectively mixed and the final Fe abundance was measured, one would find,

\begin{equation}
\mathrm{Fe_{f,2+1}}=10^{-9}\cdot \mathrm{Fe_\odot}\cdot M + 10^{-1}\cdot \mathrm{Fe_\odot}\cdot M \simeq 10^{-1}\cdot \mathrm{Fe_\odot}\cdot M.
\end{equation}

Hence, without a detailed tracking of the dynamics of the regions near the edge of the core, we can only offer the computed neutron exposures as lower bounds.

\section{GCE plots}

\label{sec:GCE plots}

\begin{figure}[ht]
\includegraphics[width=\textwidth]{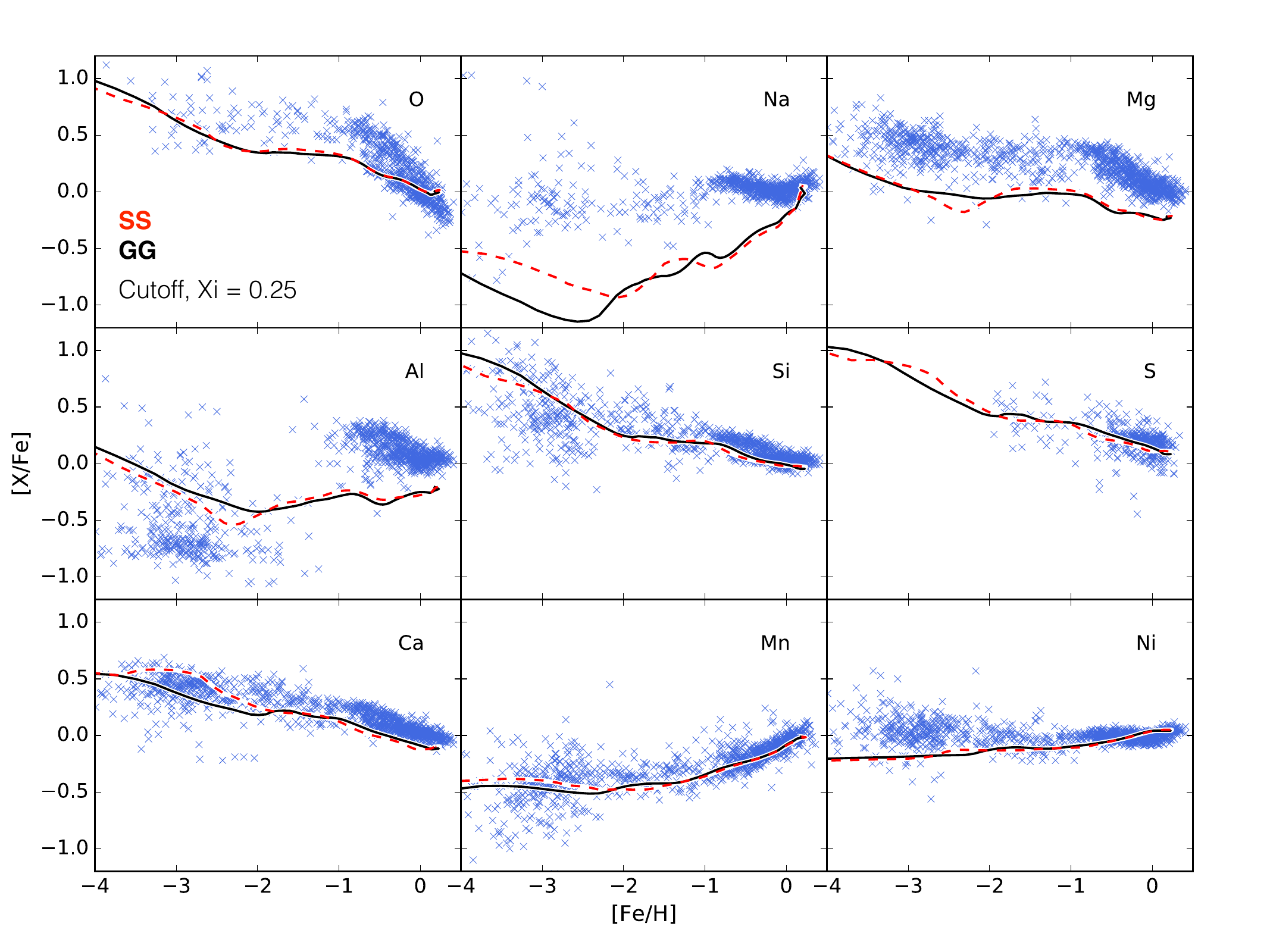}%
\caption{Selected elemental evolutions of SS and GG models using OMEGA. A compactness paramater cutoff of $\xi=0.25$ was used to assess failed /successful supernovae. Observational data is shown in blue. \label{fig:GCE1}}
\end{figure}

\begin{figure}[ht]
\includegraphics[width=\textwidth]{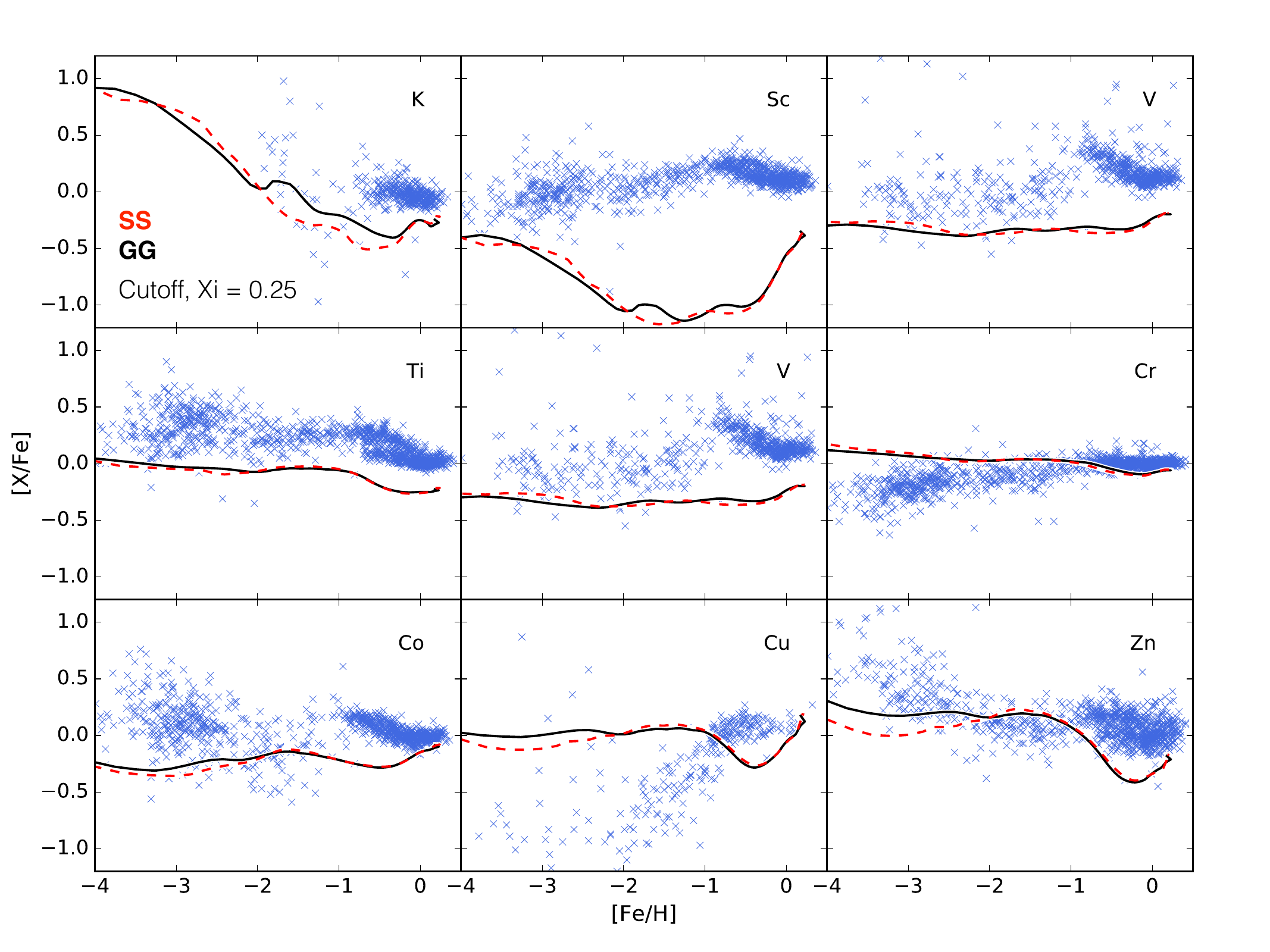}%
\caption{Selected elemental evolutions of SS and GG models using OMEGA. A compactness paramater cutoff of $\xi=0.25$ was used to assess failed /successful supernovae. Observational data is shown in blue. (cont.)\label{fig:GCE2}}
\end{figure}

\bibliography{database_2024_07_01}

\end{document}